\documentclass[twocolumn,letter]{jpsj2}
\usepackage{dcolumn}
\usepackage{bm}

\title{Superconducting Double Transition and the Upper Critical Field Limit\\of Sr$_2$RuO$_4$ in Parallel Magnetic Fields}

\author{
Kazuhiko~\textsc{Deguchi}$^{1}$\thanks{E-mail address: deguchi@scphys.kyoto-u.ac.jp}, 
Makariy~\textsc{Tanatar}$^{1,2}$\thanks{Permanent address: Inst.  Surface Chemistry, N.A.S. Ukraine.  Kyiv, Ukraine, address at present: Department of Physics, University of Toronto, Toronto, Ontario, Canada.},
Zhiqiang~\textsc{Mao}$^{1,2}$\thanks{Present address: Physics Department, Tulane University, 2001 Percival Stern, New Orleans, LA 70118.},
Takehiko~{\sc Ishiguro}$^{1,2}$,
and
Yoshiteru~\textsc{Maeno}$^{1,2,3}$
}

\inst{
$^1$Department of Physics, Graduate School of Science, Kyoto
University, Kyoto 606-8502\\
$^2$ CREST, Japan Science and Technology Corporation, Kawaguchi, 
Saitama 332-0012\\
$^3$ Kyoto University International Innovation Center, Kyoto 606-8501
}

\recdate{\today}

\def\runauthor{J. Phys. Soc. Jpn. {\bf 71}, (2002) 2839}

\abst{
We studied the specific heat and thermal conductivity of the spin-triplet superconductor Sr$_2$RuO$_4$ at low temperatures and under oriented magnetic fields $H$. We have resolved a double peak structure of the superconducting transition under magnetic field for the first time, which provides thermodynamic evidence for the existence of multiple superconducting phases. We also found a clear limiting of the upper critical field $H_{\rm c2}$ for the field direction parallel to the RuO$_2$ plane only within $\pm 2^{\circ}$. The limiting of $H_{\rm c2}$ occurs in the same $H-T$ domain of the second superconducting phase; we suggest that the two phenomena are of the same physical origin.
}

\kword{Sr$_2$RuO$_4$, spin-triplet superconductivity, superconducting double transition, upper critical field limit}

\begin{document}
\maketitle

Existence of the multiple superconducting phases is a direct proof of the complex superconducting order parameter. The well known examples include superfluid $^3$He \cite{3He-Cp-T} and the spin-triplet superconductor UPt$_3$ \cite{UPt3-Cp-T}. The superconducting state of Sr$_2$RuO$_4$ is believed to be close to these two cases \cite{Rice-old}. Since the discovery of its superconductivity \cite{discovery}, this layered ruthenate has attracted a vivid interest of physics community \cite{physicstoday}. The superconductivity of Sr$_2$RuO$_4$ has pronounced unconventional features: the invariance of spin susceptibility \cite{NMR,polarised}, appearance of spontaneous magnetic moment \cite{muSR}, absence of Hebel-Slichter peak \cite{NMR-HS}, strong dependence of $T_{\rm c}$ on nonmagnetic impurities \cite{impurity,defect}, unusual field distribution in the square flux-line-lattice \cite{SANS}, and the nodal structure of superconducting gap \cite{Cp,NQR,penetration,kappa-B}, with seemingly circular line node around a cylindrical Fermi surface \cite{kappa-L,Izawa,USA}. These features are coherently understood in terms of spin-triplet superconductivity with two-dimensional order parameter.

	Observation and characterization of multiple superconducting phases could be of notable importance for theoretical understanding of superconductivity in Sr$_2$RuO$_4$ \cite{theory}. Several recent experiments indeed suggested the existence of an additional phase boundary in magnetic fields parallel to the superconducting plane \cite{Cp,kappa-B,Mao,Yaguchi}. In this letter we give thermodynamic evidence for the second superconducting phase in Sr$_2$RuO$_4$ and present the temperature $T$ - field $H$ - inclination angle $\theta$ domain of its existence. We show that the formation of second superconducting state is closely related to the upper critical field limit in parallel field configuration, as seen in the temperature dependence of $H_{\rm c2}(T) $ and as breaking of scaling of field dependence of thermal conductivity $\kappa(H)/T$ with $H_{\rm c2}$. We argue that the multiple phases of Sr$_2$RuO$_4$ are a consequence of quasi-two-dimensionality and thus notably different from those of UPt$_3$ \cite{UPt3-Cp-T} and superfluid $^3$He \cite{3He-Cp-T}.

	The single crystals of Sr$_2$RuO$_4$ were grown by a floating-zone method in an infrared image furnace \cite{growth}. We selected the crystals with $T_{\rm c}$ above 1.4 K, close to the estimated value for impurity and defect free material ($T_{\rm c0}=1.50$ K). All the samples were characterized by ac susceptibility and x-ray measurements and were found to be of high crystalline quality, as seen in the small width of the superconducting transition (${\mathit \Delta} T_{\rm c}$) and the sharp single peaks in x-ray rocking curve measurements, with the width limited by the instrumental resolution.
\begin{figure}[b]
    \begin{center}
\includegraphics[width=7cm,clip]{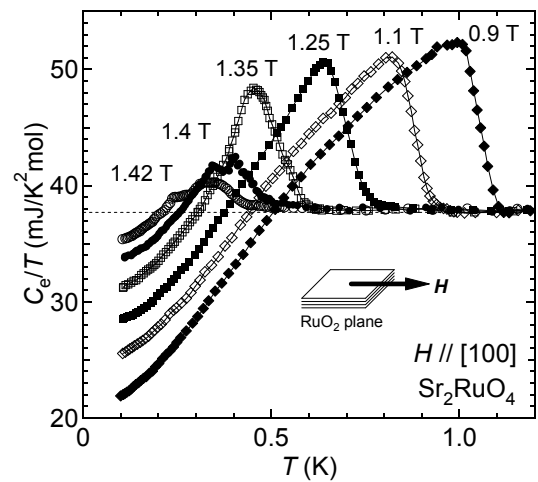}
    \end{center}
\caption{Temperature dependence of $C_{\rm e}/T$ in magnetic fields precisely parallel to the [100] direction.}
\label{fig:CDT}
\end{figure}

	For the measurements of the specific heat, the sample with $T_{\rm c} = 1.48$ K was cut and cleaved from the single crystalline rod, to a size of $2.8 \times 4.8 \times 0.50 \;{\rm mm}^{3}$. The specific heat was measured by a relaxation method with the calorimeter mounted on a single-axis sample rotator installed in a dilution refrigerator \cite{Cp}. The rotator enabled precise sample orientation with a relative accuracy of better than $0.05 ^{\circ}$. The measurements of thermal conductivity were done by one heater-two thermometers technique \cite{kappa-B}. The sample had $T_{\rm c} =1.44$ K and was mounted in a miniature vacuum cell \cite{kappa-cell}, allowing for a field alignment with an accuracy of better than $0.1 ^{\circ}$. Studies of $C_{\rm e}$ and $\kappa$ as a function of field orientation were performed in both field-cooled and zero-field-cooled states. We did not detect any difference between these sets of data beyond experimental scatter.

\begin{figure}[t]
    \begin{center}
\includegraphics[width=6.5cm,clip]{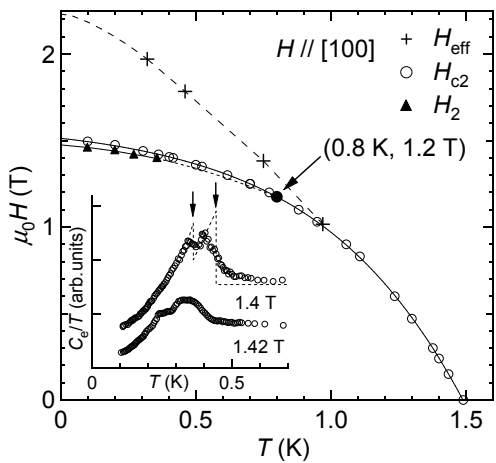}
    \end{center}
\caption{The phase diagram of Sr$_2$RuO$_4$ for $H\parallel[100]$, based on specific heat.  $H_{\rm c2}$ and $H_{\rm 2}$ are the upper critical field and critical field for the second superconducting transition. $H_{\rm eff}$ is the critical field for normalization shown in Fig.~\ref{fig:KAG}(b). The inset shows a blow-up of Fig. 1 and the definition of $H_{\rm c2}$ and $H_{\rm 2}$.}
\label{fig:HTP}
\end{figure}

	Figure~\ref{fig:CDT} shows the temperature dependence of the electronic specific heat $C_{\rm e}/T$ in magnetic field $H\parallel [100]$ direction within the conducting plane. The electronic specific heat $C_{\rm e}$ was obtained after subtraction of the phonon contribution with the Debye temperature of $410$ K. Two clearly separated specific heat jumps can be seen for $H \ge 1.4$ T in Fig.~\ref{fig:CDT} and the inset of Fig.~\ref{fig:HTP}, providing definitive evidence for a second superconducting transition. The large entropy release with a second superconducting transition suggests the transition cannot originate from the melting of the vortex lattice but the variation of the order parameter in the superconducting state. The splitting is proceeded by an anomalous feature, observed between 1.2 T and 1.4 T, where the phase transition at $T_{\rm c}(H)$ sharpens substantially, in contrast to a conventional feature observed below 1.2 T. This feature is consistent with previous reports \cite{Cp,Mao}.  In addition, the improved field alignment, the choice of appropriate field amplitude and signal-to-noise ratio enabled us to resolve the double peaks for the first time in this study.

	Figure~\ref{fig:HTP} shows the phase diagram of Sr$_2$RuO$_4$ as determined from specific heat measurements for $H\parallel [100]$ direction.  The $H_{\rm c2}$ and the critical field $H_{\rm 2}$ of the second superconducting transition are defined at the transition midpoints as in the inset of Fig.~\ref{fig:HTP}. Although the second transition feature in $C_e(T)/T$ is not observed in the field range between 1.2 T and 1.4 T, extrapolation of the $H_{\rm 2}(T)$ line to higher temperatures appears to merge with the $H_{\rm c2}(T)$ line at approximately (0.8 K, 1.2 T), where the unusual sharpening of the phase transition at $T_{\rm c}(H)$ disappears. The dotted line denoted as $H_{\rm eff}$ is based on the scaling of thermal conductivity as discussed below.
\begin{figure}[t]
    \begin{center}
\includegraphics[width=7cm,clip]{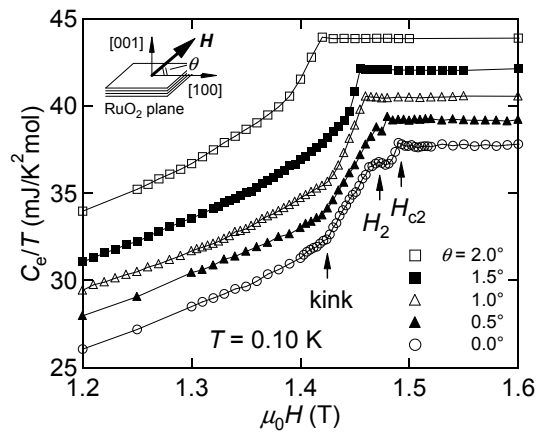}
    \end{center}
\caption{Transformation of the field dependence of $C_{\rm e}/T$ near $H_{\rm c2}$ at 0.10 K on each field angle $\theta$.  Except for $0.0^{\circ}$, each trace has an offset.}
\label{fig:CAG}
\end{figure}
\begin{figure}[b]
    \begin{center}
\includegraphics[width=7cm,clip]{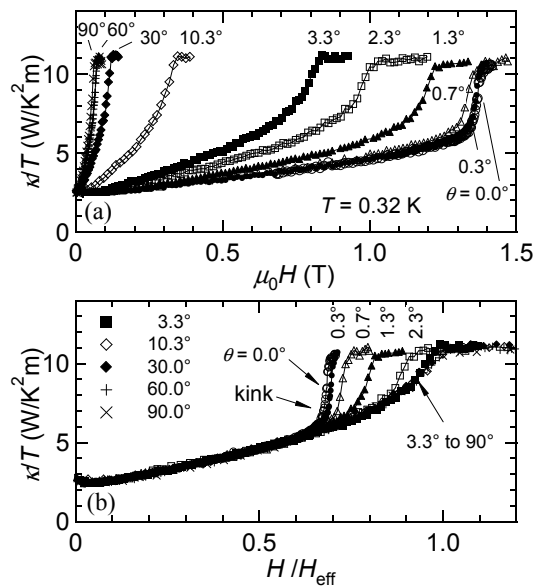}
    \end{center}
\caption{(a) Transformation of the field dependence of $\kappa/T$ at 0.32 K on each field angle $\theta$. (b) The same dependence normalized by $H_{\rm eff}$, treated as a fitting parameter.}
\label{fig:KAG}
\end{figure}

	Figure~\ref{fig:CAG} shows transformation of the field dependence of $C_{\rm e}/T$ at 0.10 K with the angle of field inclination $\theta$ from the plane. The inclination was varied within the (010) plane with $\theta = 0$ for $H\parallel [100]$. A well resolved second peak at $H_{\rm 2}$ is observed for $\theta$ from $0.0^{\circ}$ to $0.5^{\circ}$ as a shoulder on a steep increase of $C_{\rm e}/T$ near $H_{\rm c2}$. For $1.0^{\circ} \le \theta \le 2.0^{\circ}$, the structure of $H_{\rm 2}$ is no longer resolved, although a steep change of $C_{\rm e}/T$ near $H_{\rm c2}$ remains intact. Thus, very accurate alignment of the applied magnetic field to the RuO$_2$ plane is essential for inducing the second superconducting transition. The steep increase starts at the kink of $C_{\rm e}(H)/T$ and for $\theta = 0.0^{\circ}$ its magnitude corresponds to 16 \% of the electronic specific heat in the normal state $\gamma _{\rm N}$. This results shows that the destruction of the superconducting state is accompanied by an abrupt release of quasiparticles, a feature which was rarely observed before in superconductors with Fermi-liquid behavior. For $\theta > 2.0^{\circ}$, the kink vanishes and $C_{\rm e}/T$ has a smooth field dependence near $H_{\rm c2}$. This result also excludes the possibility of a mosaic structure of the crystal from the origin of the double peaks, because in that case the double peaks would not vanish with field misalignment.

	Figure~\ref{fig:KAG}(a) shows transformation of the field dependence of in-plane thermal conductivity $\kappa/T$ at 0.32 K with $\theta$. The plane of field inclination was perpendicular to the heat flow direction [010]; quite similar results were obtained for inclination within the plane parallel to the heat flow. For the in-plane magnetic field, almost half of the $\kappa(H)/T$ change between the superconducting and normal states takes place within 0.1 T below $H_{\rm c2}$. A very steep change of $\kappa(H)/T$ well corresponds to the rapid increase of $C_{\rm e}(H)/T$ slightly below $H_{\rm c2}$ in Fig.~\ref{fig:CAG}. The rapid increase is completely reversible and does not show any hysteresis with magnetic field. It has a small but finite width, and therefore is unlikely to be caused by a first order transition at $H_{\rm c2}$, as expected in the case of Pauli limiting \cite{CeCoIn5-kappa}, for example.  It is important to notice that the $\kappa(H)/T$ curves have a universal shape for $ 2^{\circ} < \theta \le 90^{\circ}$: if presented on dimensionless scale as a function of $H/H_{\rm c2}$, the curves coincide. This fact inspired our attempts to extend the scaling for the angular range $ 0^{\circ} \le \theta \le 2^{\circ}$. For this purpose we have chosen the $H_{\rm c2}$ as a fitting parameter $H_{\rm eff}$ and attempted to obtain the coincidence of all $\kappa(H)/T$ curves with the universal dependence for the fields below the kink near $H_{\rm c2}$.
\begin{figure}[b]
    \begin{center}
\includegraphics[width=7cm,clip]{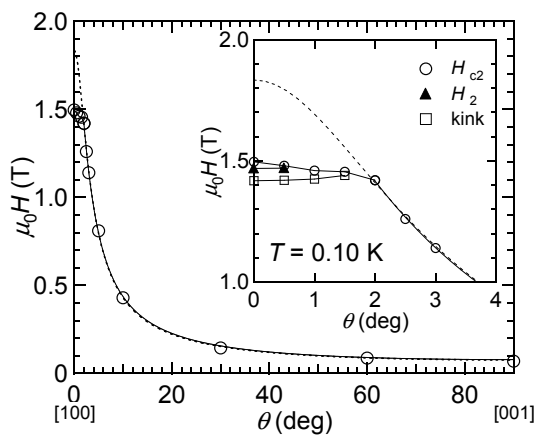}
    \end{center}
\caption{Inclination angle $\theta$ dependence of $H_{\rm c2}$, $H_{\rm 2}$ and the kink at 0.10 K. The inset shows a blow-up of the main panel. The dotted line represent fits of the Ginzburg-Landau anisotropic effective mass approximation.}
\label{fig:HAG}
\end{figure}

	The results are shown in Fig.~\ref{fig:KAG}(b), with all the curves matching below the steep increase. This result implies that in parallel field the system behaves as if the upper critical field is $H_{\rm eff}$ at low field, and the superconductivity becomes abruptly unstable on approaching the actual $H_{\rm c2}$. Thus we can conclude that there exists a mechanism which limits the upper critical field in parallel configuration and triggers immediate release of quasiparticles by sudden destruction of superconductivity. The $H_{\rm eff}(T)$ obtained in various temperatures with the same procedure for $H\parallel[100]$ is plotted in Fig.~\ref{fig:HTP}.  The deviation of $H_{\rm eff}$ from the actual $H_{\rm c2}$ is observed on condition that the temperature is below approximately 0.8 K in Fig.~\ref{fig:HTP} and the field is within $2^{\circ}$ from the RuO$_2$ plane in Fig.~\ref{fig:KAG}(b). These conditions correspond to that of the second superconducting transition. This result strongly suggests that the superconducting double transition and the limiting of upper critical field are closely linked.

	Figure~\ref{fig:HAG} shows the dependence of $H_{\rm c2}$, $H_{\rm 2}$ and the kink on the inclination angle, determined from specific heat at 0.10 K. The inset gives an expanded view of the region near $\theta = 0$. As we already mentioned, $H_{\rm 2}$ and the kink appear in close vicinity of $\theta = 0$, i.e. for $H \parallel$ RuO$_2$ plane. The region where the kink exists is slightly broader than that for $H_{\rm 2}$, with both $H_{\rm 2}$ and the kink having very weak dependence on the inclination angle. The most remarkable feature here, however, is the anomaly of $H_{\rm c2}(\theta)$. For orientations not close to $H\parallel$ RuO$_2$ plane, the angle dependence of $H_{\rm c2}$ is well fitted with Ginzburg-Landau anisotropic effective mass approximation because of interlayer coherence length $\xi_{c} = 3.2 \;{\rm nm} >$ interlayer distance $d = 0.64 \;{\rm nm}$ \cite{Tinkham}. In the region where $H_{\rm 2}$ and the kink emerge, the $H_{\rm c2}$ becomes nearly independent of $\theta$ and strongly undershoots the value obtained from extrapolation of the fitting in the $ 2^{\circ} < \theta \le 90^{\circ}$ range. This result also indicates limiting of $H_{\rm c2}$ in $H\parallel$ RuO$_2$ plane and implies that limiting of $H_{\rm c2}$ stimulates occurrence of the second superconducting transition.

	We examine possible mechanisms which can lead to the limiting of $H_{\rm c2}$ and the second superconducting transition in parallel fields. We start our discussion with the limiting of $H_{\rm c2}$. There are two known mechanism of destruction of superconductivity by the magnetic field. One is caused by increase of diamagnetic energy in superconducting state, referred to as the orbital effect. The other mechanism is caused by increase of the difference between paramagnetic energy in normal state and superconducting state, called the Pauli paramagnetic effect. This is effective in the spin-singlet state of the Cooper pair or the spin-triplet state with a field parallel to the ${\it{\textbf{d}}}$-vector. The orbital depairing is usually a much stronger effect and limits the $H_{\rm c2}$ in most cases.

	The upper critical field determined by the orbital effect varies linearly with temperature near $T_{\rm c}$, and its value at $T = 0$ is estimated from $H^{\prime}_{\rm c2}$: the slope of $H_{\rm c2}(T)$ at $T_{\rm c}$ \cite{WHH}. For a quasi two-dimensional $p$-wave superconductor in parallel field a theoretical calculation gives $H_{\rm c2}(0) = -0.75H^{\prime}_{\rm c2}T_{\rm c} = 3.3$ T \cite{Lebed}. Experimentally observed $H_{\rm c2}(0)$ by a smooth extrapolation of the low-temperature data is much lower than this value, showing that it is not limited by such orbital effect. In case of paramagnetic limiting, the transition at $H_{\rm c2}$ is believed to be of the first order. In clean systems like Sr$_2$RuO$_4$ the transition may be split into two with a new phase boundary appearing below $H_{\rm c2}(T)$ line due to formation of inhomogeneous Fulde-Ferrel-Larkin-Ovchinnikov (FFLO) superconducting state \cite{FFLO}. However, the observed sharp dependence on field orientation contradicts paramagnetic mechanisms. Furthermore, for this mechanism to be effective, the superconducting state should be spin-singlet, which is at odds with several experiments on Sr$_2$RuO$_4$ showing no change of spin susceptibility between normal and superconducting state in parallel magnetic fields. For spin-triplet superconductors there exists an orbital mechanism, which is analogous to Pauli limiting in spin-singlet superconductors \cite{Luk'yanchuk}. Here the magnetic field influences the orbital magnetic moment of the Cooper pair. The respective field is of the order of 200 T, which is notably larger than the value observed in the present experiment. From the above discussion it seems that none of the known mechanisms is satisfactory.

	Now we discuss the origin of second superconducting transition. As a possible scenario, we consider a variation of the order parameter of the superconducting state, related to internal degrees of freedom of the Cooper pairs. For Sr$_2$RuO$_4$ a lifting of degeneracy in the multi-component order parameter, $\bm{d}(\bm{k})=\hat {\bm{z}}{\mathit \Delta}_{0}(k_x + ik_y)$, in magnetic fields has been discussed by Agterberg \cite{theory}. This scenario is based on the change of orbital part of the order parameter: $\bm{d}(\bm{k})=\hat {\bm{z}}{\mathit \Delta}_{0}(k_x + ik_y) \rightarrow \hat {\bm{z}}{\mathit \Delta}_{0}k_x$. In principle, a scenario based on the variation of spin part of order parameter is also possible: $\bm{d}(\bm{k})=\hat {\bm{z}}{\mathit \Delta}_{0}(k_x + ik_y) \rightarrow (\hat {\bm{y}}+i\hat {\bm{z}}){\mathit \Delta}_{0}(k_x + ik_y)$ such as A$_{1}$-phase of superfluid $^3$He. The requirement of very precise field alignment implies the orbital scenario which is closely related to the quasi-two-dimensional Fermi surface, but even this orbital scenario has notable discrepancies with the present observation: the existence of the second phase is predicted for all temperatures below $T_{\rm c}$, $H_{\rm 2}$ is expected to be close to half of $H_{\rm c2}$, and no limiting of $H_{\rm c2}$ is expected.

	Therefore, the requirement of very precise field alignment points to the need for a new scenario based on orbital mechanism, probably specific to the quasi-two-dimensional Fermi surface and to the spin-triplet state with the orbital moment of Cooper pair perpendicular to the plane. The formation of open orbits on the cylindrical Fermi surface sheets with the parallel field configuration \cite{AMRO} may play a crucial role in the emergence of the second transition as well as in the limiting of $H_{\rm c2}$.

	In this article, we have described experimental aspects of multiple superconducting phases of Sr$_2$RuO$_4$ under in-plane magnetic fields. We have demonstrated that a clear second superconducting transition and the unusual limiting of the $H_{\rm c2}$ occur in magnetic fields precisely parallel to the RuO$_2$ plane. The two phenomena occur in the same $H-T-$inclination angle $\theta$ domain and must be closely related. These phenomena are expected to originate from the spin-triplet superconductivity combined with quasi-two-dimensionality of Sr$_2$RuO$_4$. Understanding of these phenomena will be a key to clear up the symmetry of unconventional superconductivity in Sr$_2$RuO$_4$.

	The authors would like to thank S. NishiZaki, N. Kikugawa and H. Yaguchi for their contribution to the present work. The authors also thank M. Sigrist and M. Imada for useful suggestions. This work has been in part supported by the Grant-in-Aid for Scientific Research (S) from the Japan Society for Promotion of Science (JSPS) and by the Grant-in-Aid for Scientific Research on Priority Area "Novel Quantum Phenomena in Transition Metal Oxides" from the Ministry of Education, Culture, Sports, Science and Technology (MEXT). One of the authors (K. D.) has been supported by JSPS Research Fellowship for Young Scientists.

\end{document}